\documentstyle[aps]{revtex}

\tightenlines

\begin{document}

\title{Equivalence of two thermostatistical formalisms based on the Havrda \& Charvat-Dar\'oczy-Tsallis entropies}
\author{ G.A.
Raggio\thanks{Investigador CONICET; e-mail: raggio@fis.uncor.edu}}
\address{Facultad de Matem\'atica, Astronom\'\i a y F\'\i sica \\
Universidad Nacional de C\'ordoba \\
Ciudad Universitaria, 5000 C\'ordoba, Argentina}

\date{August 14, 1999}

\maketitle

\begin{abstract} 
We show that the latest  thermostatistical formalism based on 
the Havrda \& Charvat-Dar\'oczy-Tsallis entropy $S_q [ \rho ] = (q-1)^{-1}(1- tr ( \rho^q))$ proposed by Tsallis, Mendes and Plastino is {\em equivalent} to the first one proposed by Tsallis in 1988. Here, equivalent means: {\em the ``equilibrium'' state predicted by either formalism using $q$  leads to the same expectation values for all observables as that predicted by the other formalism using $1/q$''}. We also point out once again that the basic property of {\em transitivity of equilibrium} (e.g., the $0^{th}$ Law of Thermodynamics) fails in these formalisms.

\end{abstract}

\vspace{.3in}

In a recent paper \cite{TMP}, Tsallis, Mendes and Plastino introduce a further thermostatistical formalism based on the Havrda \& Charvat-Dar\'oczy-Tsallis entropy \cite{CZECH,D,T}
\begin{equation} S_q [ \rho ] = (q-1)^{-1}(1- tr ( \rho^q) )\;\;,\;\; 0 < q \neq 1 \;.\label{entropie}\end{equation}
This third version of Tsallis' thermostatistical formalism proposes to maximize $S_q [ \rho] $ among the states $\rho$ for which
\begin{equation} <A>_{\rho}^{(3)}:= \frac{ tr (A \rho^q )}{tr (\rho^q )}  \label{3-ewert}
\end{equation}
is constant for some (operator) observable $A$. The purpose of this note is to show that this ``new'' formalism is something of a red herring: it is completely equivalent to maximizing $S_{1/q} [ \rho ]$ among the states $\rho$ for which the usual expectation value
\begin{equation} <A>_{\rho}=tr(A \rho ) \label{e-wert}
\end{equation}
is constant.

This last formalism -- indicated by the superindex $(1)$ in \cite{TMP} -- is Tsallis'  original proposal of \cite{T}; although in that paper Tsallis misidentifies the ``temperature'' in the sense that his parameter $t$ is not directly the correct ``temperature'' arising from the Legendre transformation procedure\footnote{see \cite{GR}, for a correct introduction of the ``temperature'' in this formalism along with a detailed description of some of its problems, notably the failure of the $O^{th}$-law of Thermodynamics.}. There is a further formalism - chronologically the second -- due to  Curado and Tsallis \cite{CT} (the superindex $(2)$ is used for this in \cite{TMP}) which uses the ``expectation value'' formula
\begin{equation} <A>_{\rho}^{(2)} = tr (A \rho^q ) \label{2-ewert}
\end{equation}
which has, among many other drawbacks\footnote{These are expounded in \cite{GPR}, and include the failure of the $0^{th}$-law of Thermodynamics.}, the feature that the ``expectation value'' of a sum is not the sum of the ``expectations'' and the ``expectation'' of a constant is not constant (and consequences thereof). This is remedied  in \cite{TMP} by renormalizing (\ref{2-ewert}) to get (\ref{3-ewert}). However, both these ``expectations'' fail to be affine in the state; i.e., the analogues of the relation
\begin{equation} <A>_{\lambda \rho_1+(1-\lambda)\rho_2} = \lambda <A>_{\rho_1} + (1- \lambda) <A>_{\rho_2} \label{affin}
\end{equation}
for the mixture of the states $\rho_1$ and $\rho_2$ with $0 \leq \lambda \leq 1$, fail. This relation  is crucial for a consistent probabilistic interpretation. 

We now come to our claim. First of all we are using the term ``state'' to denote a positive trace class operator of unit trace (i.e., a density operator), and ``tr'' denotes the trace. We choose to discuss these matters in the domain of quantum mechanics; the discussion in the domain of classical mechanics is identical (replace ``state'' by ``probability density'' and the trace by an appropiate integral or sum). We fix some value of $q$ greater than $0$ and not equal to $1$.

Associate to each state $\rho$ the state ${\cal R}_q (\rho )$ given by
\begin{equation} {\cal R}_q ( \rho ) := \left( tr ( \rho^q )\right)^{-1} \rho^q \;;\label{renorm}
\end{equation}
which has ${\cal R}_{1/q}$ as its inverse:
\begin{equation} {\cal R}_q ( {\cal R}_{1/q} ( \rho ))={\cal R}_{1/q} ( {\cal R}_q ( \rho ))= \rho \;;\label{invers}
\end{equation}
and clearly satisfies
\begin{equation} <A>_{\rho}^{(3)} = <A >_{{\cal R}_q (\rho )} \;.\label{bla}
\end{equation}
When $0<q <1$ and in the case of an infinite dimensional underlying Hilbert space some care has to be excercised because $tr ( \rho^q)$ and thus $S_q [ \rho]$ can be infinite. However, as shown in \cite{R}, the set of states with finite $q$-entropy is convex (and $S_q[\cdot ]$ is strictly concave for these states). In what follows we will not go into details of this and simply assume (with no loss of content or generality) that we are dealing with states of finite $q$-entropy. The following simple observation is the basis of our claim.\\

\noindent {\bf Lemma:} Let $\rho$ and $\omega$ be arbitary states, then for every $0 < q \neq 1$, one has $S_q [ \rho ] \leq S_q [ \omega ] $ if and only if $ S_{1/q}[{\cal R}_q (\rho ) ] \leq S_{1/q} [ {\cal R}_q (  \omega )]$.\\

\noindent \underline{Proof}: A simple algebraic manipulation shows that
for arbitrary states $\rho$ and $\omega$ one has:   $S_q [ \rho ] \geq S_q [ \omega ]$ if and only if $tr(\rho^q ) \leq tr ( \omega^q )$, when  $q > 1$; and  $S_q [ \rho ] \geq S_q [ \omega ]$ if and only if $tr(\rho^q ) \geq tr ( \omega^q )$, when $q > 1$. Notice also that 
\[ tr ( {\cal R}_q ( \rho )^{1/q} ) = \left( tr ( \rho^q )\right)^{-1/q} \;.\]

Let $\rho$ and $\omega$ be arbitrary states.  The inequality $tr ( \rho ^q ) \leq tr ( \omega ^q)$ is equivalent to the inequality $\left( tr ( \rho^q )\right)^{-1/q} \geq \left( tr ( \omega^q )\right)^{-1/q}$, which is equivalent to the inequality $tr \left( \left( {\cal R}_q ( \rho ) \right)^{1/q} \right) \geq tr \left( \left( {\cal R}_q ( \omega ) \right)^{1/q} \right)$. Using the  observation of the previous paragraph to shift this inequality to the entropies, one obtains the claim.\\
\\


The Tsallis-Mendes-Plastino proposal (\cite{TMP}) reads:  Maximize $S_q [ \rho ] $ among the states $\rho$ for which $<A>_{\rho}^{(3)}= \alpha $ to find  a maximizing state $\varphi_{\alpha, q}$.

My proposal would be: Maximize $S_{1/q} [ \rho ]$ among the state $\rho$ for which $<A>_{\rho} = \alpha$ to find a unique \footnote{The uniqueness, given  in \cite{GR}, is simple to prove: for given $\alpha$, if $\rho_1$ and $\rho_2$ are distinct maximizers and $S_q[\rho_1]=S_q[\rho_2]=s < \infty$, then since (\ref{e-wert}) is an affine constraint (\ref{affin}), the state $\lambda \rho_1 + (1-\lambda ) \rho_2$ has $<A>_{\lambda \rho_1 + (1-\lambda ) \rho_2}= \alpha$ for any $0 < \lambda < 1$. Since $S_q [ \cdot ]$ is strictly concave, one obtains a  contradiction via $S_q[ \lambda \rho_1 + (1-\lambda ) \rho_2] > \lambda S_q [\rho_1] + (1- \lambda )S_q[\rho_2] = s$.} maximizing state $\mu_{\alpha , 1/q}$.

Then eqs. (\ref{invers}) and (\ref{bla}) together with the  Lemma say that $\mu_{\alpha, 1/q} = {\cal R}_q ( \varphi_{\alpha, q} )$. Moreover, both proposals have the same expectation values, the computation of which never needs $\varphi_{\alpha , q}$  since these are calculated in the usual manner tracing with $\mu_{\alpha, 1/q}$. Thus, if for whatever reason, you should like to consider $S_q [\cdot ]$ as the underlying entropy then there is no need to use the strange expectation (\ref{3-ewert}) but can proceed with the usual one -- albeit replacing $q$ by $1/q$.

Explicit expressions for $\nu_{\alpha ,q}$ as well as for the ``thermodynamic functions'' associated with the map $ \alpha \mapsto S_q[ \mu_{\alpha , q}]$ via Legendre transformation are given in \cite{GR}.

I allow myself some  comments prompted by reading  \cite{TMP} as the last of many papers on the subject. I find it remarkable that a after a considerable amount of publications on the various (mainly the second) Tsallis formalisms (see the up-to-date list in the URL http://tsallis.cat.cbpf.br/biblio.htm) there is not a single piece of evidence supporting  the claim that these formalisms constitute a generalization of Boltzmann-Gibbs statistical mechanics. It is true that the formalism of statistical mechanics can be obtained by the statistical inference route (championed by Szilard and Jaynes) by maximizing the usual entropy with the usual constraint and identifying the various variables with the corresponding  thermodynamic ones. However, Boltzmann's programme (compare \cite{Lebo}) provides a way of obtaining the results of this variational excercise from the laws of physics. This link is as yet missing for the Tsallis formalisms. I also see severe problems connected with the interpretation of various unusual features of the formalisms based on $S_q$, notably the failure of the $0^{th}$ Law or more generally, the fact that the intensive parameters derived by Legendre transformation are not ``equilibrium'' parameters in the sense that they are not ``tags'' for a {\em transitive} equilibrium relation. This feature of the formalisms was pointed out in \cite{GR,GPR}.

Let me present once again the results of \cite{GR} (where the proofs are detailed) and the argument leading to this disturbing feature.  For fixed $A$ (say the energy observable) consider the
entropy function
\begin{equation}
S_q ( \alpha ) = \sup_{\{\rho \,:\; <A>_{\rho}= \alpha \} } S_q [ \rho ] \;.\label{entropiefunktion}
\end{equation}
along with its Legendre-Fenchel transform
\begin{equation}
\phi_q ( \beta ) =  \inf_{\alpha } \{ \beta \alpha - S_q (\alpha )\} = \inf_{\rho} \{ \beta <A>_{\rho} - S_q [\rho ]\} \;. \label{freie}
\end{equation}
Then $S_q ( \cdot )$ is the Legendre-Fenchel transform of $\phi_q ( \cdot )$ 
--  a  fact that really closes the circle leading to a full-fledged ``thermal'' formalism -- and there is a unique state\footnote{In the infinite dimensional situation and for $0< q < 1$, restrictions on the spectrum of $A$ have to be imposed in order to get finite solutions.} $\mu_{\alpha , q}$ such that $S_q [ \mu_{\alpha , q}] = S_q (\alpha )$; there is a unique state $\nu_{\beta , q}$ minimizing the problem (\ref{freie}); one has $\mu_{\alpha , q} = \nu_{\beta ( \alpha ), q}$ where the function $\alpha \mapsto \beta ( \alpha )$ is determined by the relation $(d \phi_q/d \beta ) ( \beta (\alpha )) = \alpha$.

The formulas for the state $\nu_{\beta , q}$ are given explicitely in \cite {GR} for the various cases: finite/infinite dimension, and $0 < q < 1$/$q>1$. 

In the most interesting case of infinite dimension with $A$ bounded below with purely discrete spectrum $\{ a_n : \; n=0,1,2, \cdots\}$, numbered in order of increasing order  not counting the multiplicities $m_n$ (assumed finite), and such that 
\[ \sum_{n \geq 1} m_n( a_n - a_o)^{q/(q-1)} < \infty \;,\]
one has
\begin{equation}
\nu{\beta , q } = \frac{ (1+(1-q)\tau_q ( \beta ) (A - a_o))^{1/(q-1)}}{\mbox{ trace of the above}} \label{q<1}
\end{equation}
for all positive\footnote{$\phi_q ( \beta )= - \infty$ for $\beta \leq 0$ so that negative ``temperatures'' do not arise.} $\beta$ when $0 < q < 1$.

For $q > 1$ and always in the infinite dimensional case with $A$ bounded below with purely discrete spectrum but no summability condition, one has
\begin{equation}
\nu{\beta , q } = \frac{ \left[ (1+(1-q)\tau_q ( \beta ) (A - a_o)\right]_{\oplus}^{1/(q-1)}}{\mbox{ trace of the above}} \label{q>1}
\end{equation}
for all $\beta$ in a certain explicitly given finite subinterval $(0, \beta_c )$  of the positive reals\footnote{$\phi_q ( \beta ) = - \infty$ for $\beta \leq 0$.}. Above $\beta_c$ the state $\nu_{\beta, q}$ is the equidistribution of ground states and independent of $\beta$\footnote{There is thus a minimal possible ``temperature''.}. In the last formula, $\left[ \cdots \right]_{\oplus}$ means ``take the positive part of the operator ...''. In both  formulas $\tau_q ( \beta )$ is a  certain implicitely determined, strictly increasing function depending on  $A$ (see \cite{GR}). Notice that $\nu_{\beta , q}$ is always degenerate since all eigenstates of $A$ above a certain energy depending on $\beta$ are not populated!

If one considers two subsystems 1, and 2 characterized by their respective Hamiltonians $A_1$ and $A_2$,  of one quantum system with the Hamiltonian $A= A_1 \otimes {\bf 1} + {\bf 1} \otimes A_2 + \lambda V$. Boltzmann-Gibbs statistical mechanics predicts
\[ \frac{ \exp ( - \beta A )}{tr(\exp ( - \beta A )) } \]
as the equilbrium state to the intensive parameter $\beta$. One has
\[  \frac{ \exp ( - \beta A )}{tr(\exp ( - \beta A )) } \rightarrow   \frac{ \exp ( - \beta A_1) }{tr_1(\exp ( - \beta A_1 )) }\otimes \frac{ \exp ( - \beta A_2)}{tr_2(\exp ( - \beta A_2 )) }\]
in the limit of weak coupling $\lambda \to 0$. This is the product-state formed with the equilibrium states of the subsystems for the same value of $\beta$. This is the  fact needed to obtain transitivity of $\beta$-equilbrium: two systems in $\beta$-equilibrium with a third are in $\beta$-equilibrium with each other. 

The reader is asked to insert $A= A_1 \otimes {\bf 1} + {\bf 1} \otimes A_2 + \lambda V$ in either of the formulas (\ref{q<1}) or (\ref{q>1}) and to see that when $\lambda \to 0$ one does not obtain a product state.

I am indebted to Pedro Pury for drawing my attention to \cite{TMP}, and to him and Gustavo Guerberoff with whom most of my work on the $S_q$-formalisms was shared  in 1995. The support of CONICET, CONICOR and SECYT-UNC is gratefully acknowledged.


\begin{thebibliography}{10}
\bibitem{TMP} C. Tsallis, R.S. Mendes and A.R. Plastino: {\sl The role of constraints within generalized nonextensive statistics}. Physica A {\bf 261}, 534-554 (1998).
\bibitem{CZECH} J. Havrda and F. Charvat: {\sl Quantification method of classification process: concept of structural $\alpha$-entropy}, Kybernetica 3, 30 (1967). I am indebted to Guy Jumaire for pointing this reference out to me in August 1996.
\bibitem{D} Z. Dar\'oczy: {\sl Generalized information functions}. Inform. Control {\bf 16}, 36-51 (1970).
\bibitem{T} C. Tsallis: {\sl Possible generalization of Boltzamnn-Gibbs Statistics}. J. Stat. Phys. {\bf 52}, 479-487 (1988).
\bibitem{GR} G. Guerberoff, and G.A. Raggio: {\sl Standard thermal statistics with $q$-entropies.}  J. Math. Phys. {\bf 37}, 1776-1789 (1996).
\bibitem{CT} E.M.F. Curado, and C. Tsallis: {\sl Generalized statistical mechanics: connection with thermodynamics}. J. Phys. A {\bf 24}, L69-L72 (1991).
\bibitem{GPR} G. Guerberoff, P.A. Pury, and G.A. Raggio: {\sl Nonstandard thermal statistics with $q$-entropies.}  J. Math. Phys. {\bf 37}, 1790-1811 (1996).
\bibitem{R} G.A. Raggio: {\sl Properties of $q$-entropies}. J. Math. Phys. {\bf 36}, 4785-4791 (1995).
\bibitem{Lebo} J.L. Lebowitz: {\sl Boltzmann's Entropy and Time's Arrow}. Physics Today {\bf 46 (9)}, 32-38 (1993).
\end{thebibliography}
\end{document}